\input lightmasses.sty
\brochureb{\sc f. j. yndur\'ain}{
\sc qcd bounds and estimates for light quark masses}{1}
\rightline{July, 31, 1997}
\rightline{NIKHEF-97-035}
\rightline{hep-ph}
\bigskip
\hrule height.5mm
\bigskip
\centerline{\titlbf Pure QCD Bounds and Estimates}
\smallskip
\centerline{\titlbf for Light Quark Masses}
\medskip
\centerrule{0.4cm}
\vskip1.2cm
\centerline{\fib F. J. Yndur\'ain,}
\medskip
\centerline{\addressfont NIKHEF-H}
\centerline{\addressfont P. O. Box 41 882}
\centerline{\addressfont 1009 DB Amsterdam, The Netherlands}
\centerline{and}
\centerline{\addressfont Departamento de F\'\i sica Te\'orica, C-XI}
\centerline{\addressfont Universidad Aut\'onoma de Madrid}
\centerline{\addressfont Canto Blanco, Madrid-28049,
 Spain\footnote*{\petit Permanent address. e-mail: fjy@delta.ft.uam.es}}
\setbox0=\vbox{\abstracttype{Abstract}We consider bounds on light quark masses 
that follow from positivity of the pseudoscalar correlator spectral function 
plus the assumption that perturbative QCD is valid for the correlator and its 
derivatives 
up to order $N$  
for momenta $t\geq\hat{t}$. We find that the bounds vary a lot depending 
on the assumed value of $\hat{t}$ and even, if 
it is too small ($\hat{t}\simeq 1.5,\,1.6,\,2.1\,\gev^2$ for respectively
 $N=0,\,1,\,2$), that there is incompatibility between 
the asumption of validity of perturbative QCD and positivity. This allows us to 
establish a criterion for the values of $\hat{t}$ admissible,
 and to get upper and lower bounds for $m_s$ and 
upper bounds for $m_d+m_u,\,m_d-m_u$. The upper bounds are not particularly 
interesting, but the lower ones are very tight; specifically 
we find
$$240\,\mev\leq m_s;\;16\,\mev\leq m_d+m_u$$
if we assume perturbative QCD to give a valid description of
 the correlator for $t\geq 2.2\,\gev^2$; or, if it only holds for 
$t\geq 4.5\,\gev^2$ then
$$150\,\mev\leq m_s;\;10\,\mev\leq m_d+m_u.$$ 
Here the masses are running masses defined at $1\,\gev$. We also show 
reasonable models where the bounds are saturated. The results suggest 
that some of the current estimates 
of the light quark masses are less precise than 
ordinarily claimed.}
\vskip2.4cm
\centerline{\box0}
\brochureendcover{Typeset with \physmatex}

\brochuresection{\S 1. Introduction}
Quark masses cannot be measured directly because of confinement, so indirect 
methods have to be devised to estimate them. In particular, light quark ($u,\,d,\,s$)
 masses can only be obtained either from lattice simulations\ref{1} or via 
QCD sum rules, which is the method that will be used in this paper. The advantage of this method 
is twofold: first, one can check, using perturbation theory, the validity of the 
calculations and, secondly, one may use very general properties of 
spectral functions to get {\sl bounds} which will only 
depend on QCD. This second feature is what will be of interest for us here.

The use of QCD sum rules for getting light quark masses, or QCD bounds, goes 
back to refs.~2, 3 and, especially, ref.~4. Since then a large number of determinations 
of the masses have been made. Among these, we may quote those in refs.~5, 6 as 
very comprehensive ones, and  refs.~7, 8, 9 as recent determinations 
employing increasingly precise QCD calculations\ref{10,11} which 
have been becoming available. Quoting from refs.~7, 9 one has
$$\eqalign{m_u+m_d=&12\pm2.5 \,\mev\cr
m_s=&171\pm 15\,\mev\cr}\equn{(1.1)}$$
and the masses refer to the $\overline{\rm MS}$ masses, defined at 1 $\gev^2$. The 
first  value comes from ref.~7, that for $m_s$ from ref.~9.

In the present paper we will contend that the {\sl errors} quoted in \equn{(1.1)}
 are excessively optimistic as indeed a large contribution to the estimates comes not 
from QCD but from low energy models. That this is so, that the errors must be 
underestimated, follows from the fact that, as we will show, bounds 
using only perturbative QCD, in a region where it should be applicable, can be 
obtained which are hardly compatible with (1.1). What is more, we will construct 
explicit, simple models incorporating perturbative QCD and 
positivity that show that the largest source of uncertainty is the 
value of the momentum at which one assumes a perturbative 
calculation to produce a good approximation. In fact, the 
 determinations of the light quark masses do indeed depend to an 
important extent  
on low energy models, and the implicit assumption of 
when the perturbative expression takes over.

\brochuresection{\S 2. Derivation of the bounds}
We will follow ref.~4 and define the correlator,
$$\Psiv^{12}_5(t)=\ii \int\dd^4x\langle{\rm vac}|T\partial^{\mu}A^{12}_{\mu}(x)
\partial^{\nu}A^{12}_{\nu}(0)^{\dag}|{\rm vac}\rangle,\equn{(2.1)}$$
$t=-q^2$ and $|{\rm vac}\rangle$ the physical vacuum. The axial current 
is
$$A^{12}_{\mu}(x)=\bar{q}_1(x)\gamma_{\mu}\gamma_5q_2(x),\equn{(2.2)}$$
and the indices 1,2 refer to quark flavours, of which we will consider various 
pairings among the $u,\,d,\,s$. The second derivative with respect 
to $t$ of the correlator $F^{12}_5(t)=\partial^2 \Psiv^{12}_5(t)/\partial t^2$
 satisfies a dispersion 
relation of the form
$$F^{12}_5(t)=\int^{\infty}_0\dd s\,\dfrac{1}{(s+t)^3}\,\dfrac{2\imag \Psiv^{12}_5(s)}{\pi}.
\equn{(2.3)}$$
Because one can write the spectral function as
$$\imag\Psiv^{12}_5(s)=\tfrac{1}{2}
\sum_{\Gammav}\left|\langle{\rm vac}|\partial^{\mu}A^{12}_{\mu}(0)|\Gammav\rangle\right|^2
(2\pi)^4\delta_4(q-p_{\Gammav}),\equn{(2.4)}$$
it follows that $\imag\Psiv^{12}_5(s)\geq 0$: it is this
 positivity that will allow us to derive 
quite general bounds.

For sufficiently large $t$ perturbative QCD is applicable to calculate $\Psiv^{12}_5(t)$. 
To leading order in $\alpha_s$ and $m_i$ we have ($N_c$=number of colours =3)
$$F^{12}_5(t)\eqsub_{\rm LO}\dfrac{N_c}{8\pi^2}\dfrac{[m_1(t)+m_2(t)]^2}{t},
\;t\gg\Lambdav^2,\equn{(2.5a)}$$
and, for the imaginary part,
$$\imag \Psiv^{12}_5(s)\eqsub_{\rm LO}\dfrac{N_c}{8\pi}s[m_1(s)+m_2(s)]^2,
\;s\gg\Lambdav^2.\equn{(2.5b)}$$
It is very important, to get tight, reliable bounds, to use 
the information contained in both equations (2.4a,b). This is 
achieved by working with the function
$$\eqalign{\varphi_{12}(t)=&\;F^{12}_5(t)-
\int_t^{\infty}\dd s\,\dfrac{1}{(s+t)^3}\,\dfrac{2\imag \Psi^{12}_5(s)}{\pi}\cr
 &=\int^t_0\dd s\,\dfrac{1}{(s+t)^3}\,\dfrac{2\imag \Psi^{12}_5(s)}{\pi}.\cr}\equn{(2.6)}$$
Using the QCD evaluation (2.5b) for the 
spectral function $\imag \Psi^{12}_5(s)$ when $s\geq t$ in (2.6) we now get 
the the perturbative QCD estimate, 
$$\varphi_{12}(t)\eqsub_{\rm LO}\dfrac{1}{4}\;\dfrac{N_c}{8\pi^2}\dfrac{[m_1(t)+m_2(t)]^2}{t}:
\equn{(2.7)}$$
note the factor $\tfrac{1}{4}$ gained with respect to 
\equn{(2.5a)}. It will also turn out that $\varphi$ is 
better than $F_5$ in that higheer order corrections are smaller for it.

Next we use the operator version of PCAC to write
$$\partial\cdot A^{12}=\sqrt{2}f_{12}M_{12}^2\phi_{12},\equn{(2.8)}$$
where $\phi_{12}$ is the field for the pseudoscalar Goldstone boson with 
decay constant $f_{12}$ and mass $M_{12}$: $\phi_{ud}=\phi_{\pi^+},\,M_{ud}=M_{\pi^+}$, 
$\phi_{us}=\phi_{K^+},\,M_{us}=M_{K^+}$, $\phi_{ds}=\phi_{K^0},\,M_{ds}=M_{K^0}$. The 
contribution of the corresponding one-particle intermediate state to (2.4) is then calculable explicitely and we get,
$$\varphi_{12}(t)=\dfrac{4f^2_{12}M^4_{12}}{(t+M^2_{12})^3}+
\dfrac{2}{\pi}\int^t_{s_0}\dd s\;\dfrac{\imag \Psiv^{12}_5(s)}{(s+t)^3};
\equn{(2.9)}$$
$s_0=M_{3\pi}^2=9M^2_{\pi}$ for $ud$ quarks, $s_0=(M_K+2M_{\pi})^2$ for the 
$(u,d)s$ states.

Combining (2.9) with (2.7) we get immediately a bound. If we 
believe that the LO expression (2.7) is a good 
description\fnote{We will suppress the indices 12 from $\varphi_{12}$ etc. 
when they are superfluous}  of $\varphi(t)$ for $t\geq t_0$ then, because $\imag \Psiv_5$ in (2.9) 
is positive we get
$$m_1(t_0)+m_2(t_0)\geq\left\{\dfrac{2^7\pi^2f_{12}^2M^4_{12}}{3}\;
\dfrac{t_0}{(t_0+M^2_{12})^3}\right\}^{\tfrac{1}{2}}.\equn{(2.10)} $$
Our task in the coming sections lies in refining (2.10) for the various quark choices.

\brochuresection{\S 3. Derivatives. Upper and lower bounds on $m_s$}
In principle it would appear that one can improve the bound in \equn{(2.10)} by 
considering quantities related to {\sl derivatives}. We will here 
consider the $\varphi^{(N)}$,

$$\eqalign{\varphi^{(0)}(t)=
\int^t_0\dd s\,\dfrac{1}{(s+t)^3}\;\dfrac{2\imag\Psiv(s)}{\pi}=
\dfrac{4f^2M^4}{(t+M^2)^3}+
\int^t_{s_0}\dd s\,\dfrac{1}{(s+t)^3}\;\dfrac{2\imag\Psiv(s)}{\pi},\cr}\equn{(3.1{\rm a})}$$
$$\eqalign{\varphi^{(1)}(t)=
3\,\int^t_0\dd s\,\dfrac{1}{(s+t)^4}\;\dfrac{2\imag\Psiv(s)}{\pi}=
\dfrac{12f^2M^4}{(t+M^2)^4}+
3\,\int^t_{s_0}\dd s\,\dfrac{1}{(s+t)^4}\;\dfrac{2\imag\Psiv(s)}{\pi},\cr}\equn{(3.1{\rm a})}$$
$$\eqalign{\varphi^{(2)}(t)=
12\,\int^t_0\dd s\,\dfrac{1}{(s+t)^5}\;\dfrac{2\imag\Psiv(s)}{\pi}=
\dfrac{48f^2M^4}{(t+M^2)^5}+
\int^t_{s_0}\dd s\,\dfrac{1}{(s+t)^5}\;\dfrac{2\imag\Psiv(s)}{\pi},\cr}\equn{(3.1{\rm a})}$$
and $\varphi^{(0)}$ coincides with $\varphi$ as defined above. 

We will not consider higher derivatives, of order $N>2$. The QCD  
NLO (next to leading order) 
 corrections to the LO result grow with $N$, as $\log N$ for $\alpha_s$ corrections, 
and as powers of $N$ for the $O({\rm mass}^{2j}/t^j)$ corrections\ref{1,4}. We may compensate 
this by taking larger values of $t$ for larger $N$, and  
 we are thus faced with a problem of 
optimization: we have to take sufficiently 
many derivatives that we get good bounds,
 but not too many that this is offset by the ensuing growth of $t$. A very 
sophisticated optimization method is described in ref.~4; here we will not 
go that far and will consider only $N=0, 1, 2$: we prefer 
to sacrifice optimality for reliability.

We start by considering the correlators 
containing the strange quark, say the quantities $\varphi^{(N)}_{us}(t)$. These  
may be calculated in QCD if $t$ is large enough. We will here keep the LO and NLO terms\fnote{
There are more terms known than the ones we use here\ref{9,11}. The $O(\alpha_s)$ 
corrections to the nonperturbative pieces are known, as is also the $O(m^6)$ term. 
these are all subleading and of the same order in $1/t$ as the 
nonpertrbative contributions we have included. Given the 
small influence of these terms, and the fact that the more important one,  
the gluon condensate, is very poorly known, we have thought it superfluous
 to include them. The NNLO correction to 
the term quadratic in $m_s$ is also known; but not that for the 
$O(m_s^4)$ one. We have preferred to keep symmetry between the 
two as, for the $s$ quark case, they are quite comparable. We have checked that the 
inclussion of this NNLO correction would not substantially alter our results.} 
in $\alpha_s$ as well as terms of relative order $m_s^2/t$, $m_s^4/t^2$, and 
the leading (in $\alpha_s$) nonperturbative contributions asssociated 
with the nonzero condensates
$$\langle{\rm vac}|:\bar{q}(0)q(0):|{\rm vac}\rangle,\;
\langle{\rm vac}|:\alpha_s G^2(0):|{\rm vac}\rangle,$$
and $q$ are the quark operators for $u,s$ quarks. The first 
condensate may be eliminated using PCAC and flavour SU(3) invariance 
in favour of products of $f_K,\,M_K$. Using the results 
of refs.~4, 10,
$$\dfrac{1}{\pi}\imag \Psiv_5^{us}(s)=\dfrac{3m^2_s(s)}{8\pi^2}\left\{
\left[1+\dfrac{17\alpha_s}{3\pi}\right]s-
2m_s^2(s)\left[1+\dfrac{16\alpha_s}{3\pi}\right]\right\},$$
$$
\eqalign{F_5^{us}(t)=\dfrac{3}{8\pi^2}\left\{\dfrac{m_s^2(t)}{t}
\left[1+\dfrac{11\alpha_s}{3\pi}\right]-
\dfrac{2m^4_s(t)}{t^2}\left[1+\dfrac{28\alpha_s}{3\pi}\right]\right.\cr
\left.+\dfrac{1}{t^3}\left[\dfrac{8\pi^2f^2_KM^2_K}{3}+
\dfrac{2\pi\langle\alpha_s:G^2:\rangle}{3}\right]\right\}}$$
and we then have,
$$\eqalign{\varphi_{us}^{(0)}(t)=\dfrac{3}{8\pi^2}
\left\{\dfrac{m_s^2}{t}
\left[\tfrac{1}{4}+\left(\tfrac{5}{12}+2\log 2\right)\dfrac{\alpha_s}{\pi}\right]
-\dfrac{2m_s^4}{t^2}\left[\tfrac{3}{4}+\left(6+4\log 2\right)\dfrac{\alpha_s}{\pi}\right]\right.\cr
\left. +\dfrac{1}{t^3}\left[\dfrac{8\pi^2f_K^2M_K^2}{3}+
\dfrac{2\pi\langle\alpha_s:G^2:\rangle}{3}\right]\right\};\cr
}\equn{(3.2{\rm a})}$$
$$\eqalign{\varphi_{us}^{(1)}(t)=\dfrac{3}{8\pi^2}
\left\{\dfrac{m_s^2}{t^2}
\left[\tfrac{1}{2}+\left(\tfrac{7}{3}+2\log 2\right)\dfrac{\alpha_s}{\pi}\right]
-\dfrac{4m_s^4}{t^3}\left[\tfrac{7}{8}+\left(\tfrac{49}{6}+4\log 2\right)\dfrac{\alpha_s}{\pi}\right]
\right.\cr
\left.+\dfrac{3}{t^4}\left[8\pi^2f_K^2M_K^2+2\pi\langle\alpha_s:G^2:\rangle\right]\right\};
}\equn{(3.2{\rm b})}$$
$$\eqalign{\varphi_{us}^{(2)}(t)=\dfrac{3}{8\pi^2}
\left\{\dfrac{m_s^2}{t^3}
\left[\tfrac{11}{16}+\left(\tfrac{187}{48}+\log 2\right)\dfrac{\alpha_s}{\pi}\right]
-\dfrac{12m_s^4}{t^4}\left[\tfrac{15}{16}+\left(\tfrac{29}{3}+4\log 2\right)\dfrac{\alpha_s}{\pi}
\right]\right.\cr
\left.+\dfrac{12}{t^5}\left[8\pi^2f_K^2M_K^2+2\pi\langle\alpha_s:G^2:\rangle\right]\right\}
}\equn{(3.2{\rm c})}$$

with $m=m_s(t)$ the two loop runnuing mass\ref{12}
$$\eqalign{m(t)=\hat{m}\left(\tfrac{1}{2}\log t/\Lambdav^2\right)^{-d_m}
\left[1-d_1\dfrac{\log\log t/\Lambdav^2}{\log t/\Lambdav^2}+
d_2\dfrac{1}{\log t/\Lambdav^2}\right];\cr
d_m=\dfrac{4}{\beta_0},\;d_1=8\,\dfrac{51-\tfrac{19}{3}n_f}{\beta_0^3},\;d_2=\dfrac{8}{b_0^3}
\Big[\left(\tfrac{101}{12}-\tfrac{5}{18}n_f\right)\beta_0
-51+\tfrac{19}{3}n_f\Big], \cr
\beta_0=11-\tfrac{2}{3}n_f.\cr}\equn{(3.3)}$$
An extra advantage of using the combination giving the $\varphi^{(N)}(t)$, and 
not simply the derivatives $\partial^NF_5(t)/\partial t^N$, is that, as announced, 
and as Eqs.~(3.2) show, 
the NLO corrections for the first are substantially smaller 
than for the last. This will make the calculations 
based on perturbative QCD more reliable.

As discussed, we may combine Eqs.~(3.1), (3.2) to derive bounds on $m_s$. 
We do so and rewrite the result as
$$Am_s^4-Bm_s^2+C\leq 0\equn{(3.4)}$$
where $A, B, C$ may be read from (3.1,2); for $\varphi^{(0)}$ and to LO,
$$A=\dfrac{N_c}{8\pi^2}\,\dfrac{6}{4t^2},\;B=\dfrac{N_c}{8\pi^2}\,\dfrac{1}{4t},
\;C=\dfrac{4f_K^2M_K^4}{(t+M_K^2)^3}$$
and corresponding expressions to NLO. Obviously, \equn{(3.4)} only has a solution if 
$$B^2\geq 4AC\equn{(3.5)}$$
so, unless this condition is satisfied we find an incompatibility  
between the validity of the QCD expression and positivity. Defining the {\sl critical} values 
 $t_N$ those for which we get equality in \equn{(3.5)}, and the corresponding equations 
obtained from the various terms in Eqs.~(3.2), we get the values
$$\matrix{\nada&{\rm LO}&{\rm NLO}&{\rm NLO+NP}\cr
t_0=&1.76\;\gev^2&1.47\;\gev^2&1.36\;\gev^2\cr
t_1=&2.09\;\gev^2&1.75\;\gev^2&1.61\;\gev^2\cr
t_2=&2.46\;\gev^2&2.22\;\gev^2&2.04\;\gev^2.\cr}\equn{(3.6)}$$
The tag ``NP" indicates that we 
have included the nonperturbative pieces as in Eqs.~(3.2);
we have taken the following values for the parameters
$$M_K=495.7\;\mev,\;f_K=115\;\mev,\;\Lambdav(n_f=3, 1\; {\rm loop})=300\;\mev.$$
It is interesting to note that the values of the $t_N$ do not change much from LO to 
NLO to NLO+NP and that they generally {\sl decrease} from LO to NLO to NLO +NP, as would
 be expected: one imagines 
that the NLO expression is valid for smaller values of $t$ than the LO one, 
and also that including NP effects improves the convergence.
The values of $t_N$ found are rather large: comparable to the lower range among 
those employed in the 
calculations of e.g. refs.~7, 8, 9.
If we assumed that the perturbative QCD expression would 
be valid as soon as $t=t_N$ (something 
that cannot be the case) we would have obtained evaluations of $m_s$. These 
give too large values, such that $m_s(1\;\gev^2)\simeq 450\;\mev$. What we {\sl do} is to 
consider that the QCD 
evaluation of the $N$/th derivative at $t$ 
is to be trusted if the QCD evaluation of the {\sl next} derivative $N+1$ 
is compatible with positivity. Thus, at LO we can use the zeroth derivative if $t\geq t_1=
2.09\;\gev^2$, and the first one if $t\geq t_2=2.46\;\gev^2$. We 
will refer to the bounds so obtained as optimum, or optimist bounds. 
Alternatively, we may want to play it safe and require $t\geq2 t_N$. 
The corresponding bounds are reported in Table I, where we have also included 
the bounds obtained for $N=2\;{\rm and}\;t=6.5\;\gev^2$.
\setbox1=\vbox{\hsize=0.8\hsize
\smallskip
$$\eqalign{
N=0;\;\matrix{\nada&{\rm LO}&{\rm NLO}&{\rm NLO+NP}\cr
t_1&295<m_s<596&240<m_s<477&245<m_s<458\cr
2t_1&158<m_s<1018&142<m_s<847&150<m_s<803\cr
}\cr
N=1;\;\matrix{t_2&308<m_s<609&234<m_s<527&243<m_s<506\cr
2t_2&162<m_s<1048&140<m_s<916&148<m_s<867\cr
}\cr
N=2;\,\matrix{t=6.5;&156<m_s<1140&129<m_s<1038&128<m_s<1052\cr}\cr}$$
\smallskip
 \centerrule{40mm}
\smallskip
\centerline{\petit Table I. Bounds, in \mev, for various
 values of $t$ on $m_s\equiv m_s(1\;\gev^2)$.}}
\medskip
\centerline{\boxit{\box1}}
\medskip
The bounds are stable 
from LO to NLO to NLO+NP, and also for 
the various values of $N$. 
The upper bound is not very interesting, as it is well above all existing estimates; 
but the lower one is very tight, as, indeed, it is violated by several of the calculations 
found in the literature (refs.~1, 8, 9) and is barely compatible with others\ref{5,6}. 
We will discuss this in a latter section. For the moment we
 summarize the lower bounds in the two possibilities,
$$\matrix{\hbox{Optim. bound:}\;&244\;\mev<m_s\cr
\hbox{Safe bound:}\;&149\;\mev<m_s.\cr}\equn{(3.7)}$$

\brochuresection{\S 4. Lower bounds on $m_d\pm m_u$}
\brochuresubsection{4.1. Lower bound on $m_d+ m_u$ }
The combination $ud$ allows us to derive bounds on $m_d+m_u$. The equations 
are similar to (3.1, 2) with the replacement $m_s\rightarrow m_d+m_u$,
 $f_K\rightarrow f_{\pi}=95\;\mev$ and $M_K\rightarrow M_{\pi}=137.3\;\mev$ and, 
moreover, taking $n_f=2,\,\Lambdav=350\,\mev$. The 
term in $m^4$ is slightly  different, but 
it is utterly negligible now: for this reason only 
lower bounds may be obtained. We choose the values of the $t$ 
at which we calculate the bounds to be the same as those for the $s$ quark case. Specifically, 
we define throughout this section $t_1=1.75\,\gev^2,\;t_2=2.22\,\gev^2$. Then we 
have the results in Table II,
\setbox1=\vbox{\hsize=0.7\hsize
\smallskip
$$
\matrix{{\rm NLO},\cr\;N=0\cr}\;\Bigg\{\matrix{t_1=1.75\;\gev^2;&16.3<m_d+m_u\cr
2t_1=3.5\;\gev^2;&9.8<m_d+m_u\cr}$$
\smallskip
$$
\matrix{{\rm NLO},\cr\;N=1\cr}\;\Big\{\matrix{t_2=2.2\;\gev^2;&16.7<m_d+m_u\cr
2t_2=4.5\;\gev^2;&9.8<m_d+m_u\cr}$$
\smallskip
$${\rm NLO},\;N=2;\; t=6.5\;{\gev}^2;\; 9.0<m_d+m_u$$
\smallskip 
\centerrule{38mm}
\smallskip
\centerline{\petit Table II. NLO bounds in, \mev, 
for $m_d+m_u\equiv m_d(1\;\gev^2)+m_u(1\;\gev^2)$.}}
\medskip
\centerline{\boxit{\box1}}
\medskip
The bounds are also now 
very stable. We may, as for the $s$ quark case, highlight
the best bounds,
$$\matrix{\hbox{Optim. bound:}\;&16.5\;\mev<m_d+m_u\cr
\hbox{Safe bound:}\;&9.8\;\mev<m_d+m_u.\cr}\equn{(4.1)}$$
It may perhaps be remarked that the bounds for the $m_d+m_u$ combination are more 
reliable than for the $m_s$ case. This 
is because the NP contributions 
are much smaller 
in this case.  At the values of $t$ we are considering of just a few percent; and 
the same is true of $O(m^6)$ ones, 
still smaller. 
\brochuresubsection{4.2. Lower bound on $m_d-m_u$.}
We now consider the {\sl difference} between $\varphi_{ds}$ and $\varphi_{us}$. We will 
still be able to calculate this from QCD, but a rigorous 
proof of the positivity of 
$$\delta\Psiv(s)\equiv\Psiv_5^{ds}(s)-\Psiv_5^{us}$$
is not possible. However, it is very likely that this 
positivity holds; in fact, it follows in the chiral SU(2) limit provided 
one assumes that $(m_d-m_u)/m_d\gg (m_d,m_u)/m_s$, an inequality 
that is amply satisfied in all estimates. Moreover, the positivity of $\delta\Psiv(s)$ 
may be checked experimentally on the Kaon pole, and from QCD at large $s$. So 
we will assume it.

A second problem now is that we have to subtract, from the pole terms, 
the electromagnetic contrubutions to the mass differences, as they are 
comparable to the masses themselves. We will use for this the chiral 
dynamics estimate\ref{13}
$$M_{K^0}^2-M_{K^+}^2=(1.9\pm0.5)M_{\pi^0}^2-M_{\pi^+}^2$$
so we replace, in the difference of the pole terms,
$$M_{K^0}^4-M_{K^+}^4\rightarrow \delta M_K^4\equiv (M_{K^0}^4-M_{K^+}^4)_{\rm physical}-
(M_{K^0}^4-M_{K^+}^4)_{\rm e.m.}=(5.18\pm0.48)M_K^4.\equn{(4.2)}$$
We will furthermore assume that $f_{K^+}=f_{K^0}$. We thus have, from QCD and using 
the NLO calculations of refs.~4, 10,
$$\eqalign{\delta \varphi^{(0)}(t)=\dfrac{2N_c\Deltav}{8\pi^2t}
\left\{\left[\tfrac{1}{4}+
\left(\tfrac{5}{12}+2\log 2\right)\dfrac{\alpha_s}{\pi}\right]
-\dfrac{m_s^2}{t}\,\left[\tfrac{3}{4}+
\left(\tfrac{1}{2}+4\log 2\right)\dfrac{\alpha_s}{\pi}\right]\right\},\cr}
\equn{(4.3{\rm a})}$$
$$\eqalign{\delta \varphi^{(1)}(t)=\dfrac{6N_c\Deltav}{8\pi^2t^2}
\left\{\left[\tfrac{1}{6}+
\left(\tfrac{7}{9}+\tfrac{2}{3}\log 2\right)\dfrac{\alpha_s}{\pi}\right]
-\dfrac{m_s^2}{t}\,\left[\tfrac{7}{12}+
\left(\tfrac{7}{8}+\tfrac{8}{3}\log 2\right)\dfrac{\alpha_s}{\pi}\right]\right\},\cr}
\equn{(4.3{\rm b})}$$
$$\Deltav=m_s(t)\left[m_d(t)-m_u(t)\right]$$
on one hand, and on the other
$$\eqalign{\delta \varphi^{(0)}(t)=\dfrac{4f_K^2\delta M_K^4}{(t+M_K^2)^3}+({\rm positive})\cr
\delta \varphi^{(1)}(t)=\dfrac{12f_K^2\delta M_K^4}{(t+M_K^2)^4}+({\rm positive}).\cr}
\equn{(4.4)}$$
We now get two types of bounds. If we make no assumptions on $m_s$, we 
still obtain lower bounds on $m_d-m_u$; but 
of course much better bounds are obtained if assuming a reasonable value 
for the strange quark mass. The results are given in Table III.
\setbox1=\vbox{\hsize=0.7\hsize
\smallskip
$$
\matrix{{\rm NLO},\cr\;N=0\cr}\;\Bigg\{\matrix{t_1:\;6.22\pm0.6<m_d-m_u\cr
2t_1:\;2.57\pm 0.3<m_d-m_u\cr}$$
\smallskip
$$\matrix{{\rm NLO},\cr\;N=1\cr}\;\Bigg\{\matrix{t_2:\;6.32\pm0.6<m_d-m_u\cr
2t_2:\;2.52\pm 0.3<m_d-m_u\cr}$$
\smallskip 
\centerrule{20mm}
\smallskip
\centerline{\petit Table III. Bounds, in \mev, 
for $m_d-m_u\equiv m_d(1\;\gev^2)-m_u(1\;\gev^2)$.}
\centerline{\petit $m_s(1\;\gev^2)=200\;\mev$.}
}
\medskip
\centerline{\boxit{\box1}}
\medskip
The bounds are also here very stable. If we leave $m_s$ as a free parameter 
the bounds deteriorate and we find,
$$N=0;\;t_1:\;m_d-m_u>3.37\,\mev;\;N=1;\;t_2:\;m_d-m_u>3.33\,\mev\equn{(4.5)}$$
and they are obtained with $m_s\sim 530\,\mev$. For the choice 
$2t_1,\,2t_2$ the bounds decrease to $1\;\mev$ and are attained with 
$m_s\sim800\,\mev$.
The bounds for $m_s=200\;\mev$ are very thight in the sense that they essentially coincide 
with the existing {\sl estimates}\ref{5,6}. This poses the problem of the errors, and 
corrections to the bounds, to which we now turn.

\brochuresection{\S 5. Errors and corrections to the bounds. Estimates of masses}
It is not the purpose of this paper to present a 
new evaluation of light quark masses; but we want to give at least estimates of how much 
the bounds may be expected to deviate from the true values of theese quantities. 
We will give the detailed calculations for $m_d+m_u$, for which 
the methods are more 
reliable, and at the 
end present the  results corresponding to $m_s$. Also we will consider the case $N=1$,
assuming perturbative QCD to be valid above $\hat{t}$, 
with $\hat{t}=t_2=2.2\,\gev^2$ and $\hat{t}=2t_2=4.5\,\gev^2$.

Let us rewrite the 
equations for clarity of reference. From (3.1b) and the equation for $ud$ 
analogous to (3.2b) we have, equating the QCD expression 
and the dispersive representation,
$$\dfrac{[m_d(t)+m_u(t)]^2}{2t^2}=\dfrac{12f_{\pi}M^4_{\pi}}{(t+M_{\pi}^2)^4}
+\dfrac{6}{\pi}\int^t_{s_0}\dd s\;\dfrac{\imag \Psiv_5(s)}{(s+t)^4}\;
,s_0=9M_{\pi}^2.\equn{(5.1)}$$
(We have written the LO expression, but NLO evaluations
 will be performed throughout this section).
If $\imag \Psiv_5(s)$ vanished in the interval $s_0\leq s\leq \hat{t}$ then, 
by putting $t=\hat{t}$ in \equn{(5.1)} 
the lower bounds would become equalities. So, to determine how 
tight are the bounds, and to estimate $m_d+m_u$ we require models 
for $\imag \Psiv_5(s)$ in that low momentum  region.

In the lower end of the interval we may use 
chiral dynamics to evaluate the contribution of the $3\pi$ intermediate 
state. The calculation is elementary and one finds\ref{5}
$$\imag \Psiv_5(s)=\dfrac{M_{\pi}^2s}{768\pi^3f^2_{\pi}},\equn{(5.2)}$$
an approximation that we expect to be valid until the opening of the $\rho\,\pi$ threshold, 
at $s=s_{\rho\pi}=(M_{\rho}+M_{\pi})^2$. The contribution of (5.2) is minute, and will 
consequently be neglected. From $s_{\rho\pi}$ onwards 
we expect that the continuum of $\imag \Psiv_5(s)$ will be dominated by 
the $\rho\pi$ intermediate state as happens e.g. in $e^+e^-\rightarrow\hbox{hadrons}$ 
annihilations. One could estimate the contribution 
of this channel with the help of vector meson dominance in the soft 
limit, which is certainly not a very accurate model. Since we 
are only interested in an estimate we will merely interpolate 
between zero at $s_{\rho\pi}$ and 
the QCD perturbative value at $\hat{t}$. Thus we 
consider $\imag \Psiv_5^{\rm continuum}\simeq\imag\Psiv_5^{\rho\pi}$ and
$$\dfrac{1}{\pi}\imag\Psiv_5^{\rho\pi}(s)=\cases{0,s\leq s_{\rho\pi}\cr
\dfrac{3}{8\pi^2}\,\dfrac{[m_d(\hat{t})+m_u(\hat{t})]^2}{\hat{t}-s_{\rho\pi}}s(s-s_{\rho\pi})
\left[1+\dfrac{17\alpha_s(\hat{t})}{3\pi}\right],\, s_{\rho\pi}\leq s\leq \hat{t}}\equn{(5.3)}$$
so that $\imag\Psiv_5^{\rho\pi}(\hat{t})=\imag\Psiv_5^{\rm pert.\,QCD}(\hat{t})$.

Besides the $\rho\pi$ continuum we have the 
contribution of the $\pi'=\pi(1300)$ resonance. We may write this as
$$\dfrac{6}{\pi}\imag\Psiv_5^{\pi'}(s)=12rf^2_{\pi}M_{\pi}^4\delta(s-M^2_{\pi'}).\equn{(5.4)}$$
The quantity $r$, ratio between the wave functions at the 
origin of the $\bar{u}d$ in the $\pi$, $\pi'$, is not 
known, nor can it be obtained in any direct manner from the $\pi'$ width. 
In a constituent quark model, $r\sim 0.3\;{\rm to}\;0.5$; and 
similar values are obtained in bag models.
  In a purely Coulombic model one would get 
$r=1/8$. So we allow $r$ to vary 
in the range $0\leq r\leq 0.5$. Anyway, the contribution of the $\pi'$ is 
rather small.

With all this we write
$$\dfrac{6}{\pi}\int^t_{s_0}\dd s\,\dfrac{\imag\Psiv_5(s)}{(s+t)^4}
\simeq\dfrac{12f^2_{\pi}M^4_{\pi}}{(t+M^2_{\pi'})^4}\,r+
\dfrac{6}{\pi}\int^t_{s_{\rho\pi}}\dd s\,\dfrac{\imag\Psiv_5^{\rho\pi}(s)}{(s+t)^4}.\equn{(5.5)}$$
Plugging this into (5.1) with $t=\hat{t}=t_2,\,2t_2$ we get the 
estimates reported in Table IV.
\setbox1=\vbox{\hsize=0.65\hsize
\smallskip
$$\matrix{\nada&{\rm bound}&r=0&r=0.25&r=0.5\cr
\hat{t}=t_2=2.2\,\gev^2&16.7&19.4&19.6&19.9\cr
\hat{t}=2t_2=4.5\,\gev^2&9.8&12.5&12.9&13.3\cr}$$
\smallskip
 \centerrule{40mm}
\smallskip
\centerline{\petit Table IV. Bounds and estimates, in \mev, for $m_d+m_u$.}}
\medskip
\centerline{\boxit{\box1}}
\medskip
As is seen here the main source 
of error (and a {\sl large} error it is) comes from the variation of the region $t\geq\hat{t}$ 
where we believe that perturbative QCD may be applied to evaluate $\imag\Psiv_5$. On 
this one has no control. For  $e^+e^-\rightarrow\hbox{hadrons}$ we know 
that the cross section is well described 
by perturbative QCD from 
$s\sim 1\;{\rm to}\;2\;\gev^2$. If we assume this, then $\hat{t}\sim 2\,\gev^2$ and 
 $m_d+m_u\sim 20\,\mev$. 
But one may argue that the NLO correction in  $e^+e^-\rightarrow\hbox{hadrons}$ is 
small, $\alpha_s/\pi$, while that for 
$\imag\Psiv_5$ is large, $17\alpha_s/3\pi$. This suggests a latter onset 
of the perturbative regime for the latter quantity, say at $2t_2\sim4.5\;\gev^2$ and then 
 $m_d+m_u\sim13\,\mev$.

For $m_s$ similar considerations would apply and we 
get the results summarized in Table V.
\setbox1=\vbox{\hsize=0.65\hsize
\smallskip
$$\matrix{\nada&{\rm bound}&r=0&r=0.25&r=0.5\cr
\hat{t}=t_2=2.2\,\gev^2&221&242&250&254\cr
\hat{t}=2t_2=4.5\,\gev^2&138&169&175&180\cr}$$
\smallskip
 \centerrule{40mm}
\smallskip
\centerline{\petit Table V. Bounds and estimates, in \mev, for $m_s$.}}
\medskip
\centerline{\boxit{\box1}}
\medskip
The estimates take into account NLO and NP corrections. The 
effective two-body threshold is now given by $K\rho$ or $K^*\pi$, 
with an average mass squared of $1.3\,\gev^2$.  
For the mass of the
 $K'$ resonance we have guessed $M_{K'}=1.5\,\gev$. Like for $m_d+m_u$ 
the resonance contributes little, while a very large 
variation occurs when we move the region where 
the onset of the perturbative regime takes place. also in common with the 
$ud$ case we find that the bound we had  obtained are rather tight.

Before finishing this section a few words have to be said 
to clarify further the meaning of the results reported in Tables IV, V 
and say a few words comparing them with other derivations. As 
for the first, we note that our results 
were obtained by comparing the QCD expression, say for $N=0$ and the 
$ud$ case $F_5^{\rm QCD}(t)$, and the dispersive integral, that we may call $F_5^{\rm disp.}(t)$, 
obtained from the model to $s=t$ and QCD above it:
$$\eqalign{F_5^{\rm disp.}(t)\equiv \dfrac{4f_{\pi}^2M_{\pi}^4}{(t+M_{\pi}^2)^3}+
r\dfrac{4f_{\pi}^2M_{\pi}^4}{(t+M_{\pi'}^2)^3}\cr
+\dfrac{2}{\pi}
\int^t_{ s_{\rho\pi}}\dd s\,\dfrac{\imag\Psiv_5^{\rho\pi}(s)}{(s+t)^3}+
\dfrac{2}{\pi}
\int^{\infty}_t\dd  s\,\dfrac{\imag\Psiv_5^{\rm QCD}(s)}{(s+t)^3}.
\cr}$$ 
 The estimates of the Tables correspond 
to requiring equality at $t=\hat{t}$; but of course what is 
really needed is equality, up to neglected higher order 
corrections (in our evaluation, $O(\alpha_s^2)$), for {\sl all} $t\geq \hat{t}$. Equlity as 
$t\rightarrow \infty$ is guaranteed; so we expect that we will have it for 
intermediate $t$ as well. We have checked this numerically for $N=0$, the $ud$ case 
with 
$$\hat{t}=2.2\,{\gev}^2,\;r=0.5,\;m_d+m_u=19.9\,\mev.$$
Here we get, for the ratios $\rho_0(t)=F_5^{\rm QCD}(t)/F_5^{\rm disp.}(t)$, 
$\rho_1(t)={F'}_5^{\rm QCD}(t)/{F'}_5^{\rm disp.}(t)$ the values
$$\matrix{&t=2.2&4.5&8.5&30&100&\,\gev^2\cr
\rho_0=&1.22&1.10&1.04&1.01&1.005&\cr
\rho_1=&1.06&1.03&0.96&0.93&0.94&\cr}$$
i.e., what one would expect\fnote{The fact that the largest error occurs, for 
$\rho_0(t)$, at  
$t=\hat{t}$ where nominally one should have equality is 
easily understood if we realize that, to determine the values 
of $m_d+m_u$, we have used \equn{(3.2a)}, which only coincides with 
$F_5(t)$ minus the integral $2\int_t^{\infty}\dd s\imag\Psiv_5^{\rm QCD}(s)/\pi(s+t)^3$ to 
corrections $O(1/\log^2t/\Lambdav^2)$.} for a calculation with an error $O(\alpha_s^2)$. Actually, 
virtual equality may be obtained if we 
replace the rather crude model for $\imag\Psiv_5^{\rho\pi}$, with a linear threshold (\equn{(5.3)})
 by a more realistic model, e.g., with a square-root threshold for the  
$\rho\pi$ channel.

Lastly, our evaluations show why some other estimates find such small  
values for the masses. Specifically, considering ref.~9, probably the more complete (from 
the point of view of perturbative QCD) evaluation of $m_s$ we see that the very low 
value for this quantity, $m_s=171\,\mev$, is obtained because the authors there 
assume that perturbative QCD holds for $F_5(t)$ for $t\sim 2,\,3\,\gev^2$; but a model 
is employed for $\imag\Psiv_5(s)$ which is well {\sl below} the 
peturbative QCD value up to very large momenta, $s\sim6.5\,\gev^2$. 
If, in the same calculation, we took the perturbative value for $\imag\Psiv_5(s)$ 
down to $2.2\;{\rm to}\;4.5\,\gev^2$ we 
would get values quite compatible with those 
reported in Table V, $175\;{\rm to}\;250\,\mev$.      

\brochuresection{\S 6. Summary and discussion}
The results of this note may be viewed as, first, a set of bounds such that, to go 
below them would imply that perturbative QCD fails at unreasonably low values of $t$, say 
$t\sim 4\;{\to}\;5\,\gev^2$: these are the ``Safe" bounds of Eqs.~(3.7), (4.1). 
Secondly,  we find  
indications that the actual values may easily be larger (the ``Optim." bounds 
in the same equations). The evaluations of the last section, Tables IV and V, 
are to be viewed as ``existence proofs" that the 
bounds can be saturated using reasonable physical assumptions. Putting all together 
we may draw the conclusion that current estimations of the {\sl errors} 
in the evaluations of quark masses are excessively optimistic. We would consider 
brackets
$$\eqalign{140\,\mev\leq m_s\leq 254\,\mev,\cr
10\,\mev\leq m_d+m_u\leq 20\,\mev}\equn{(6.1)}$$
to represent realistic, {\sl attainable} estimates. For the 
difference $m_d-m_u$ the bounds are good only if 
we restrict $m_s$. If we assume this to be bounded 
as in \equn{(6.1)}, then we find
$$2\,\mev\leq m_d-m_u \leq 11\,\mev\equn{(6.2)}$$
to be a generous, but attainable, bracket.  

One may wonder how high one has to put the onset of 
the perturbative regime 
if we take values for the quark masses as low as those in some recent latice and other 
determinations\ref{1,14}. The answer is, impossibly high. Considering for 
example the $s$ quark, even if we assume 
that perturbative QCD is only valid at $t\geq\hat{t}= 6.5\,\gev^2$,
 and only the use of $\varphi^{(0)}$
 is allowed (no derivatives) one still has $m_s>110\,\mev$, 
and, for $\hat{t}= 10\,\gev^2$, $m_s>90\,\mev$. This 
assuming that the spectral function {\sl vanishes} completely below 
the corresponding values of $t$: if we include estimates for 
the low momentum piece like those in \S 4 we increase the values 
of $m_s$ by at least 25\%. For the combination $m_d+m_u$ the corresponding bounds are 
$8$ and $7$ \mev.

We want finally to say a few words about bounds on the individual $u,d$ masses, and 
the connections among the various determinations. In the present 
work we have only used perturbative QCD and 
positivity; but more information may be 
obtained using chiral dynamics evaluations that 
permit estimates of {\sl ratios} of quark 
masses. Using this one may disentangle relations like (6.1, 2). This is 
what is done in a recent paper by Lelouch, de Rafael and Taron\ref{15}, in which 
questions similar to the ones raised here are also discussed. This paper, which appeared 
after the present work was finished, is largely complementary to ours: as stated, 
chiral dynamics estimations of the mass ratios are included, but the 
question of the compatibility of perturbative QCD 
with positivity is not raised.

\brochuresection{Acknowledgement}
We would like to thank NIKHEF, where most of this work was done, for hospitality and 
financial support.

\vfill\eject
\brochuresection{References}
\item{1}{{\sc A.Gonz\'alez-Arroyo,
  G. Martinelli and F. J. Yndur\'ain, \jpl{117}{1982}{437}; 
R. Gupta and T. Bhattacharya, \jpr{D55}{1977}{7203}; C. R. Allton et al., 
\jnp{421}{1994}{667}; B. J. Gough et al.,} preprint hep-ph/9610223, 1996.}
\item{2}{\sc F. J. Yndur\'ain, \jpl{63}{1976}{211.}}
\item{3}{\sc V. A. Novikov et al., {\sl Sov. J. Nucl. Phys.}, {\bf 27} (1978) 2, 274.}
\item{4}{\sc C. Becchi S. Narison, E. de Rafael and F. J. Yndur\'ain, \jzp{C8}{1981}{335.}}
\item{5}{\sc J. Gasser and H. Leutwyler, {\sl Phys. Rep.} {\bf C87} (182) 77.}
\item{6}{\sc C. A. Dom\'\i nguez and E. de Rafael,
 {\sl Ann. Phys. } (NY), {\bf 174} (1987) 372.}
\item{7}{\sc J. Bijnens, J. Prades and E. de Rafael, \jpl{348}{1995}{226.}}
\item{8}{\sc M. Jamin and M. M\"untz, \jzp{C66}{1995}{633.}}
\item{9}{\sc K. G. Chetyrkin et al., \jpr{D51}{1995}{5090.}}
\item{10}{\sc D. J. Broadhurst, \jpl{101}{1981}{423.}}
\item{11}{\sc S. C. Generalis, {\sl J. Phys.}, {\bf G16} (1990) 787; L. R. Sugurladze, 
and F. V. Tkachov, \jnp{B331}{1990}{35};
 K. G. Chetyrkin, S. G. Groshny and F. V. Tkachov, \jpl{119}{1982}{407}; 
S. G. Groshny, A. L. Kataev, S. A. Larin and L. R. Sugurladze, 
\jpr{D43}{1991}{1633}; P. Pascual and E. de Rafael, \jzp{12}{182}{127}.}
\item{12}{\sc R. Tarrach, \jnp{183}{1981}{384}.}
\item{13}{\sc J. Bijnens, \jpl{306}{1993}{343}; 
J. F. Donoghue, B. R. Holstein and D. Wyler, \jpr{D47}{1993}{2089}.}
\item{14}{\sc D. B. Kaplan and A. V. Manohar \jprl{56}{1986}{2004}.}
\item{15}{{\sc L Lelouch, E. de Rafael and J. Taron}, Marseille preprint 
CPT-97/P.3519, hep-ph/9707523, 1997.}

\bye